\PassOptionsToPackage{unicode}{hyperref}
\PassOptionsToPackage{hyphens}{url}
\PassOptionsToPackage{dvipsnames,svgnames,x11names}{xcolor}
\documentclass[
]{article}
\usepackage{amsmath,amssymb}
\usepackage{lmodern}
\usepackage{iftex}
\ifPDFTeX
  \usepackage[T1]{fontenc}
  \usepackage[utf8]{inputenc}
  \usepackage{textcomp} 
\else 
  \usepackage{unicode-math}
  \defaultfontfeatures{Scale=MatchLowercase}
  \defaultfontfeatures[\rmfamily]{Ligatures=TeX,Scale=1}
\fi
\IfFileExists{upquote.sty}{\usepackage{upquote}}{}
\IfFileExists{microtype.sty}{
  \usepackage[]{microtype}
  \UseMicrotypeSet[protrusion]{basicmath} 
}{}
\makeatletter
\@ifundefined{KOMAClassName}{
  \IfFileExists{parskip.sty}{%
    \usepackage{parskip}
  }{
    \setlength{\parindent}{0pt}
    \setlength{\parskip}{6pt plus 2pt minus 1pt}}
}{
  \KOMAoptions{parskip=half}}
\makeatother
\usepackage{xcolor}
\usepackage{graphicx}
\makeatletter
\def\maxwidth{\ifdim\Gin@nat@width>\linewidth\linewidth\else\Gin@nat@width\fi}
\def\maxheight{\ifdim\Gin@nat@height>\textheight\textheight\else\Gin@nat@height\fi}
\makeatother
\setkeys{Gin}{width=\maxwidth,height=\maxheight,keepaspectratio}
\makeatletter
\def\fps@figure{htbp}
\makeatother
\NewDocumentCommand\citeproctext{}{}
\NewDocumentCommand\citeproc{mm}{%
  \begingroup\def\citeproctext{#2}\cite{#1}\endgroup}
\makeatletter
 \let\@cite@ofmt\@firstofone
 \def\@biblabel#1{}
 \def\@cite#1#2{{#1\if@tempswa , #2\fi}}
\makeatother
\newlength{\cslhangindent}
\newlength{\csllabelwidth}
\newenvironment{CSLReferences}[2] 
 {\begin{list}{}{%
  \setlength{\itemindent}{0pt}
  \setlength{\leftmargin}{0pt}
  \setlength{\parsep}{0pt}
  \ifodd #1
   \setlength{\leftmargin}{\cslhangindent}
   \setlength{\itemindent}{-1\cslhangindent}
  \fi
  \setlength{\itemsep}{#2\baselineskip}}}
 {\end{list}}
\usepackage{calc}

\ifLuaTeX
\usepackage[bidi=basic]{babel}
\else
\usepackage[bidi=default]{babel}
\fi
\babelprovide[main,import]{american}

\def\languageshorthands#1{}
\ifLuaTeX
  \usepackage{selnolig}  
\fi
\IfFileExists{bookmark.sty}{\usepackage{bookmark}}{\usepackage{hyperref}}
\IfFileExists{xurl.sty}{\usepackage{xurl}}{} 
\hypersetup{
  pdftitle={Caustics: A Python Package for Accelerated Strong
Gravitational Lensing Simulations},
  pdfauthor={Connor Stone, Alexandre Adam, Adam Coogan, M. J.
Yantovski-Barth, Andreas Filipp, Landung Setiawan, Cordero Core, Ronan
Legin, Charles Wilson, Gabriel Missael Barco, Yashar Hezaveh, Laurence
Perreault-Levasseur},
  pdflang={en-US},
  colorlinks=true,
  linkcolor={Maroon},
  filecolor={Maroon},
  citecolor={Blue},
  urlcolor={Blue},
  pdfcreator={LaTeX via pandoc}}

\title{Caustics: A Python Package for Accelerated Strong Gravitational
Lensing Simulations}

\usepackage[affil-it]{authblk}
\usepackage{orcidlink}
\author[1,2,3%
  *%
  \ensuremath\mathparagraph]{Connor Stone%
    \,\orcidlink{0000-0002-9086-6398}\,%
    }
\author[1,2,3%
  *%
  ]{Alexandre Adam%
    \,\orcidlink{0000-0001-8806-7936}\,%
    }
\author[1,2,3%
  *%
  ]{Adam Coogan%
    \,\orcidlink{0000-0002-0055-1780}\,%
      \thanks{Work done while at UdeM, Ciela, and Mila}%
  }
\author[1,2,3%
  ]{M. J. Yantovski-Barth%
    \,\orcidlink{0000-0001-5200-4095}\,%
    }
\author[1,2,3%
  ]{Andreas Filipp%
    \,\orcidlink{0000-0003-4701-3469}\,%
    }
\author[5%
  ]{Landung Setiawan%
    \,\orcidlink{0000-0002-1624-2667}\,%
    }
\author[5%
  ]{Cordero Core%
    \,\orcidlink{0000-0002-3531-3221}\,%
    }
\author[1,2,3%
  ]{Ronan Legin%
    \,\orcidlink{0000-0001-9459-6316}\,%
    }
\author[1,2,3%
  ]{Charles Wilson%
    \,\orcidlink{0000-0001-7071-5528}\,%
    }
\author[1,2,3%
  ]{Gabriel Missael Barco%
    \,\orcidlink{0009-0008-5839-5937}\,%
    }
\author[1,2,3,4%
  ]{Yashar Hezaveh%
    \,\orcidlink{0000-0002-8669-5733}\,%
    }
\author[1,2,3,4%
  ]{Laurence Perreault-Levasseur%
    \,\orcidlink{0000-0003-3544-3939}\,%
    }

\affil[1]{Ciela Institute - Montréal Institute for Astrophysical Data
Analysis and Machine Learning, Montréal, Québec, Canada}
\affil[2]{Department of Physics, Université de Montréal, Montréal,
Québec, Canada}
\affil[3]{Mila - Québec Artificial Intelligence Institute, Montréal,
Québec, Canada}
\affil[4]{Center for Computational Astrophysics, Flatiron Institute, 162
5th Avenue, 10010, New York, NY, USA}
\affil[5]{eScience Institute Scientific Software Engineering Center,
1410 NE Campus Pkwy, Seattle, WA 98195, USA}
\affil[$\mathparagraph$]{Corresponding author: %
}
\affil[*]{These authors contributed equally.}
\date{19 March 2024}

\begin{document}
\maketitle

\section{Summary}\label{summary}

Gravitational lensing is the deflection of light rays due to the gravity
of intervening masses. This phenomenon is observed in a variety of
scales and configurations, involving any non-uniform mass such as
planets, stars, galaxies, clusters of galaxies, and even the large scale
structure of the universe. Strong lensing occurs when the distortions
are significant and multiple images of the background source are
observed. The lens objects must align on the sky of order \(\sim 1\)
arcsecond for galaxy-galaxy lensing, or 10's of arcseonds for
cluster-galaxy lensing. As the discovery of lens systems has grown to
the low thousands, these systems have become pivotal for precision
measurements and addressing critical questions in astrophysics. Notably,
they facilitate the measurement of the Universe's expansion rate (e.g.
\citeproc{ref-holycow}{K. C. Wong et al., 2020}), dark matter (e.g.
\citeproc{ref-Hezaveh2016}{Hezaveh et al., 2016};
\citeproc{ref-Vegetti2014}{Vegetti \& Vogelsberger, 2014}), supernovae
(e.g. \citeproc{ref-Rodney2021}{Rodney et al., 2021}), quasars (e.g.
\citeproc{ref-Peng2006}{Peng et al., 2006}), and the first stars (e.g.
\citeproc{ref-Welch2022}{Welch et al., 2022}) among other topics. With
future surveys expected to discover hundreds of thousands of lensing
systems, the modelling and simulation of such systems must occur at
orders of magnitude larger scale then ever before. Here we present
\texttt{caustics}, a Python package designed to handle the extensive
computational demands of modeling such a vast number of lensing systems.

\section{Statement of need}\label{statement-of-need}

The next generation of astronomical surveys, such as the Large Synoptic
Sky Telescope survey and Euclid wide survey, are expected to uncover
hundreds of thousands of gravitational lenses
(\citeproc{ref-Collett2015}{Collett, 2015}), dramatically increasing the
scientific potential of gravitational lensing studies. Currently,
analyzing a single lensing system can take several days or weeks, which
will be infeasible as the number of known lenses increases by orders of
magnitude. Thus, advancements such as computational acceleration via
GPUs and/or algorithmic advances such as automatic differentiation are
needed to reduce the analysis timescales. Machine learning will be
critical to achieve the necessary speed to process these lenses, it will
also be needed to meet the complexity of strong lens modelling.
\texttt{caustics} is built with the future of lensing in mind, using
\texttt{PyTorch} (\citeproc{ref-pytorch}{Paszke et al., 2019}) to
accelerate the low level computation and enable deep learning algorithms
which rely on automatic differentiation. Automatic differentiation also
benefits classical algorithms such as Hamiltonian Monte Carlo
(\citeproc{ref-hmc}{Betancourt, 2017}). With these tools available,
\texttt{caustics} will provide greater than two orders of magnitude
acceleration to most standard operations, enabling previously
impractical analyses at scale.

Several other simulation packages for strong gravitational lensing are
already publicly available. The well established \texttt{lenstronomy}
package has been in use since 2018 (\citeproc{ref-lenstronomy}{Birrer et
al., 2021}); \texttt{GLAMER} is a C++ based code for modelling complex
and large dynamic range fields (\citeproc{ref-GLAMER}{Metcalf \&
Petkova, 2014}); \texttt{PyAutoLens} is also widely used
(\citeproc{ref-PyAutoLens}{Nightingale et al., 2021}); and GIGA-Lens is
a specialized JAX (\citeproc{ref-JAX}{Bradbury et al., 2018}) based
gravitational lensing package (\citeproc{ref-GIGALens}{Gu et al.,
2022}); among others (\citeproc{ref-Galan2021}{Galan et al., 2021};
\citeproc{ref-Keeton2011}{Keeton, 2011}; \citeproc{ref-Kneib2011}{Kneib
et al., 2011}; \citeproc{ref-Wagner2024}{Wagner-Carena et al., 2024}).
There are also several ``in house'' codes developed for specialized
analysis which are then not publicly released (e.g.
\citeproc{ref-Suyu2010}{Suyu \& Halkola, 2010}). \texttt{Caustics}
development has been primarily focused on three aspects: processing
speed, user experience, and flexibility. The code is optimized to fully
exploit PyTorch's capabilities, significantly enhancing processing
speed. The user experience is streamlined by providing three interfaces
to the code: configuration file, object-oriented, and functional, where
each interface level requires more expertise but allows more
capabilities. In this way, users with all levels of gravitational
lensing simulation experience may effectively engage with the software.
Flexibility is achieved by a determined focus on minimalism in the core
functionality of \texttt{caustics}. All of these elements combine to
make \texttt{Caustics} a capable lensing simulator to support machine
learning applications, and classical analysis.

\texttt{Caustics} fulfills a timely need for a differentiable lensing
simulator. Several other fields have already benefitted from such
simulators, for example: gravitational wave analysis
(\citeproc{ref-Coogan2022}{Coogan et al., 2022};
\citeproc{ref-Edwards2023}{Edwards et al., 2023};
\citeproc{ref-Wong2023}{K. W. K. Wong et al., 2023}); astronomical image
photometry (\citeproc{ref-Stone2023}{Stone et al., 2023}); point spread
function modelling (\citeproc{ref-Desdoigts2023}{Desdoigts et al.,
2023}); light curves (\citeproc{ref-Millon2024}{Millon et al., 2024});
and even generic optimization for scientific problems
(\citeproc{ref-Nikolic2018}{Nikolic, 2018}). \texttt{Caustics} is built
on lessons from other differentiable codes, with the goal of enabling
machine learning techniques in the field of strong lensing. With
\texttt{caustics} it will now be possible to analyze over 100,000 lenses
in a timely manner (\citeproc{ref-Hezaveh2017}{Hezaveh et al., 2017};
\citeproc{ref-Perreault2017}{Perreault Levasseur et al., 2017}).

\section{Scope}\label{scope}

\texttt{Caustics} is a gravitational lensing simulator. The purpose of
the project is to streamline the simulation of strong gravitational
lensing effects on the light of a background source. This includes a
variety of parametric lensing profiles such as: Singular Isothermal
Ellipsoid (SIE), Elliptical Power Law (EPL), Pseudo-Jaffe,
Navarro-Frenk-White (NFW), and External Shear. Additionally, it offers
non-parametric representations such as a gridded convergence or
potential field. For the background source \texttt{caustics} provides a
Sérsic light profile, as well as a pixelized light image. Users may
easily extend these lens and source lists using templates provided in
the documentation.

Once a lensing system has been defined, \texttt{caustics} may then
perform various computational operations on the system such as
raytracing through the lensing system, either forwards and backwards.
Users may compute the lensing potential, convergence, deflection field,
time delay field, and magnification. All of these operations may readily
be performed in a multi-plane setting to account for interlopers or
multiple sources. Since the code is differentiable (via PyTorch), one
may easily also compute the derivatives of these quantities, such as
when finding critical curves or computing the Jacobian of the lens
equation. For example, one may project interlopers from multiple lensing
planes to a single plane by computing the effective convergence,
obtained with the trace of the Jacobian.

\begin{figure}
\centering
\includegraphics{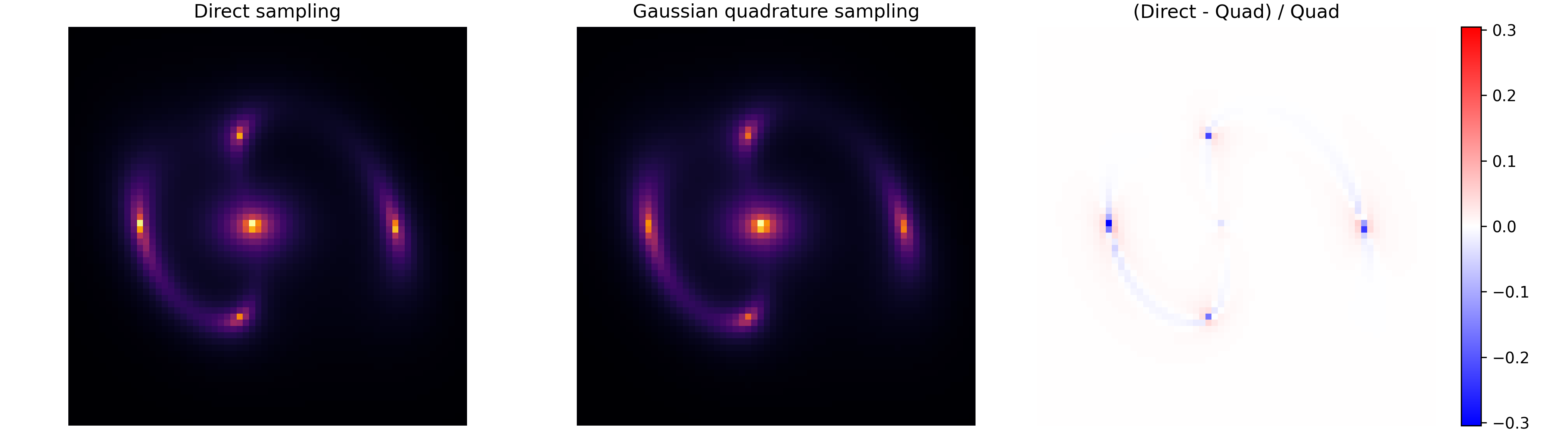}
\caption{Example simulated gravitational lens system defined by a Sérsic
source, SIE lens mass, and Sérsic lens light. Left, the pixel map is
sampled only at the midpoint of each pixel. Middle, the pixel map is
supersampled and then integrated using gaussian quadrature integration
for greater accuracy. Right, the fractional difference between the two
is shown. We can see that in this case the midpoint sampling is
inaccurate by up to 30\% of the pixel value in areas of high contrast.
The exact inaccuracy depends greatly on the exact
configuration.\label{fig:sample}}
\end{figure}

With these building blocks in place, one may construct fast and accurate
simulators used to produce training sets for machine learning models or
for inference on real-world systems. Neural networks have become a
widespread tool for amortized inference of gravitational lensing
parameter (\citeproc{ref-Hezaveh2017}{Hezaveh et al., 2017}) or in the
detection of gravitational lenses (\citeproc{ref-Huang2021}{Huang et
al., 2021}; \citeproc{ref-Petrillo2017}{Petrillo et al., 2017}), but
they require large and accurate training sets which can be created
quickly with \texttt{caustics}. A demonstration of such a simulator is
given in \autoref{fig:sample} which also demonstrates the importance of
sub-pixel sampling. This involves raytracing through the lensing mass
and extracting the brightness of the background source. Further, the
image then must be convolved with a PSF for extra realism. All of these
operations are collected into a single simulator which users may access
and use simply as a function of the relevant lensing and light source
parameters. Since \texttt{caustics} is written in \texttt{PyTorch}, its
simulators are differentiable, thus one can compute gradients through
the forward model. This enables machine learning algorithms such as
recurrent inference machines (\citeproc{ref-Adam2023}{Adam et al.,
2023}) and diffusion models (\citeproc{ref-Adam2022}{Adam et al.,
2022}).

The current scope of \texttt{caustics} does not include weak lensing,
microlensing, or cluster scale lensing simulations. While the underlying
mathematical frameworks are similar, the specific techniques commonly
used in these areas are not yet implemented, though they represent an
avenue for future development.

\texttt{Caustics}' defined scope ends at lensing simulation, thus it
does not include functionality to optimize or sample the resulting
functions. Users are encouraged to use already existing optimization and
sampling codes like \texttt{scipy.optimize}
(\citeproc{ref-scipy}{Virtanen et al., 2020}), \texttt{emcee}
(\citeproc{ref-emcee}{Foreman-Mackey et al., 2013}), \texttt{dynesty}
(\citeproc{ref-dynesty}{Speagle, 2020}), \texttt{Pyro}
(\citeproc{ref-pyro}{Bingham et al., 2019}), and \texttt{torch.optim}
(\citeproc{ref-pytorch}{Paszke et al., 2019}).

\section{Performance}\label{performance}

\begin{figure}
\centering
\includegraphics{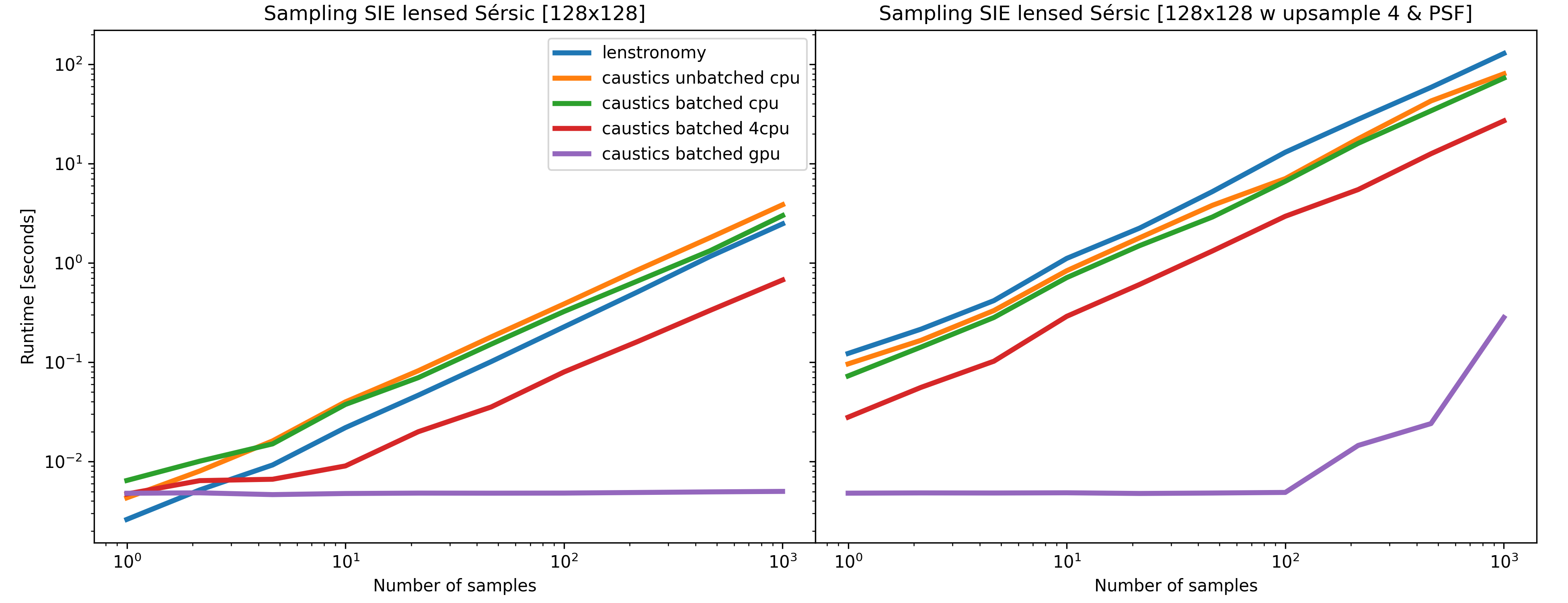}
\caption{Runtime comparisons for a simple lensing setup. We compare the
amount of time taken (y-axis) to generate a certain number of lensing
realizations (x-axis) where a Sérsic model is lensed by an SIE mass
distribution. For CPU calculations we use
\texttt{Intel\ Gold\ 6148\ Skylake} and for the GPU we use a
\texttt{NVIDIA\ V100}, all tests were done at 64 bit precision. On the
left, the lensing system is sampled 128 pixel resolution only at pixel
midpoints. On the right, a more realistic simulation includes upsampled
pixels and PSF convolution. From the two tests we see varying
performance enhancements from compiled, unbatched, batched,
multi-threaded, and GPU processing setups.\label{fig:runtime}}
\end{figure}

Here we discuss the performance enhancements enabled by
\texttt{caustics}. Via \texttt{PyTorch}, the code allows operations to
be batched, multi-threaded on CPUs, or offloaded to GPUs to optimize
computational efficiency. In \autoref{fig:runtime} we demonstrate this
by sampling images of a Sérsic with an SIE model lensing the image (much
like \autoref{fig:sample}). In the two subfigures we show performance
for simply sampling a 128x128 image using the pixel midpoint (left), and
sampling a ``realistic'' image (right) which is upsampled by a factor of
4 and convolved with a PSF. All parameters are randomly resampled for
each simulation to avoid caching effects. This demonstrates a number of
interesting facts about numerical performance in such scenarios.

We compare the performance with that of \texttt{Lenstronomy} as our
baseline. The most direct comparison between the two codes can be
observed by comparing the \texttt{Lenstronomy} line with the ``caustics
unbatched cpu'' line. \texttt{Lenstronomy} is written using the
\texttt{numba} (\citeproc{ref-numba}{Lam et al., 2015}) package which
compiles python code into lower level C code. The left plot shows that
\texttt{caustics} suffers from a significant overhead compared with
\texttt{Lenstronomy}, which is nearly twice as fast as the ``caustics
unbatched cpu'' line. This occurs because the pure Python (intepreted
language) elements of \texttt{caustics} are much slower than the C/Cuda
PyTorch backends (compiled language). This is most pronounced when fewer
computations are needed to perform a simulation. Despite this overhead,
\texttt{caustics} showcases a strong performance when using the batched
GPU setting, especially in the more realistic scenario with extra
computations in the simulator including 4x oversampling of the
raytracing and the PSF convolution.

Comparing the ``caustics unbatched cpu'' and ``caustics batched cpu''
lines we see that batching can provide more efficient use of the same,
single CPU, computational resources. However, in the realistic scenario
the batching has minimal performance enhancement, likely because the
python overhead of a for-loop is no longer significant compared to the
large number of numerical operations being performed.

Comparing ``caustics batched cpu'' and ``caustics batched 4cpu'' we see
that PyTorch's automatic multi-threading capabilities can indeed provide
performance enhancements. However, the enhancement is not a direct
multiple of the number of CPUs due to overhead. For tasks that are
``embarrassingly parallel,'' such as running multiple MCMC chains, it is
more effective to parallelize at the job level rather than at the thread
level to avoid these overheads.

The most dramatic improvements are observed when comparing any CPU
operations with ``caustics batched gpu''. Although communication between
the CPU and GPU can be slow, consolidating calculations into fewer,
larger batches allows caustics to fully exploit GPU capabilities. In the
midpoint sampling, the GPU never ``saturates'' meaning that it runs
equally fast for any number of samples. In the realistic scenario we
reach the saturation limit of the GPU memory at 100 samples and could no
longer simultaneously model all the systems, we thus entered a linear
regime in runtime just like the CPU sampling does for any number of
simulations. Nonetheless, it is possible to easily achieve over 100X
speedup over CPU performance, making GPUs by far the most efficient
method to perform large lensing computations such as running many MCMC
chains or sampling many lensing realizations (e.g.~for training machine
learning models).

\section{User experience}\label{user-experience}

Caustics offers a tiered interface system designed to cater to users
with varying levels of expertise in gravitational lensing simulation.
This section outlines the three levels of interfaces that enhance user
experience by providing different degrees of complexity and flexibility.

\textbf{Configuration file interface:} The most accessible level of
interaction is through configuration files. Users can define simulators
in \texttt{.yaml} format, specifying parameters such as lens models,
light source characteristics, and image processing details like PSF
convolution and sub-pixel integration. The user may then load such a
simulator in a single line of Python and carry on using that simulator
as a pure function \texttt{f(x)} which takes in parameters such as the
Sérsic index, position, SIE Einstein radius, etc. and returns an image.
This interface is straightforward for new users and for simplifying the
sharing of simulation configurations between users.

\textbf{Object oriented interface:} This intermediate level allows users
to manipulate lenses and light sources as objects. The user may build
simulators just like the configuration file interface, or they may
interact with the objects in a number of other ways accessing further
details about each lens. Each lensing object has (where meaningful) a
convergence, potential, time delay, and deflection field. We provide
examples to visualize all of these. Users may apply the full flexibility
of Python with these lensing objects and construct customized analysis
code, though there are many default routines which enable one to quickly
perform typical analysis tasks.

\begin{figure}
\centering
\includegraphics{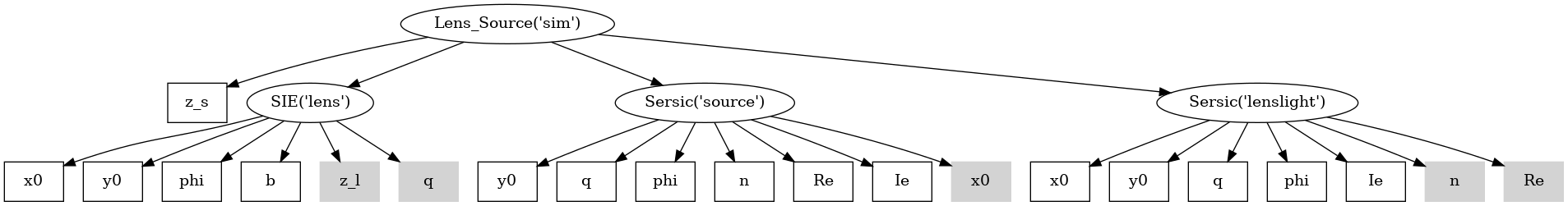}
\caption{Example directed acyclic graph representation of the simulator
from \autoref{fig:runtime}. Ellipses are \texttt{caustics} objects and
squares are parameters; open squares are dynamic parameters and greyed
squares are static parameters. Parameters are passed at the top level
node (\texttt{Lens\_Source}) and flow down the graph automatically to
all other objects which require parameter values to complete a lensing
simulation.\label{fig:graph}}
\end{figure}

For both the object oriented and \texttt{.yaml} interfaces, the final
simulator object can be analyzed in a number of ways,
\autoref{fig:graph} demonstrates how one can investigate the structure
of a simulator in the form of a directed acyclic graph of calculations.
Note that one may also fix a subset of parameter values, making them
``static'' instead of the default which is ``dynamic''. Users can
produce such a graph representation in a single line of Python for any
\texttt{caustics} simulator.

\textbf{Functional interface:} The functional interface eschews the
object oriented \texttt{caustics} code, instead giving the user access
to individual mathematical operations related to lensing, most of which
are drawn directly from gravitational lensing literature. All such
functions include references in their documentation to the relevant
papers and equation numbers from which they are derived. These equations
have been tested and implemented in a reasonably efficient manner. This
interface is ideal for researchers and developers looking to experiment
with novel lensing techniques or to modify existing algorithms while
leveraging robust, pre-tested components.

Each layer is in fact built on the one below it, making the transition
from one to the other a matter of following documentation and code
references. This makes the transition easy since one may very clearly
observe how their current analysis can be reproduced in the lower level.
\texttt{Caustics} thus provides a straightforward pipeline for users to
move from beginner to expert. Users at all levels are encouraged to
investigate the documentation as the code includes extensive docstrings
for all functions, including units for most functions. This transparency
not only aids in understanding and utilizing the functions correctly but
also enhances the reliability and educational value of the software.

\section{Flexibility}\label{flexibility}

The flexibility of caustics is fundamentally linked to its design
philosophy, which is focused on providing a robust yet adaptable
framework for gravitational lensing simulations. A focus on minimalism
in the core functionality means that research-ready analysis routines
must be built by the user. To facilitate this, our Jupyter notebook
tutorials include examples of many typical analysis tasks, with the
details laid out for the user so they may simply copy and modify to suit
their particular analysis task. Thus, we achieve flexibility both by
allowing many analysis paradigms, and by supporting the easy development
of production code.

Research is an inherently dynamic process and gravitational lensing is
an evolving field. Designing flexible codes for such environments
ensures long lasting relevance. \texttt{Caustics} is well poised to grow
and evolve with the needs of the community.

\section{Machine Learning}\label{machine-learning}

One of the core purposes of \texttt{caustics} is to advance the
application of machine learning to strong gravitational lensing
analysis. This is accomplished through two avenues. First, as
demonstrated in \autoref{fig:runtime}, \texttt{caustics} efficiently
generates large samples of simulated mock lensing images by leveraging
GPUs. Since many machine learning algorithms are ``data hungry'', this
translates to better performance with more examples to learn from.
Literature on machine learning applications in strong gravitational
lensing underscores the benefits of this generation capacity
(\citeproc{ref-Brehmer2019}{Brehmer et al., 2019};
\citeproc{ref-Chianese2020}{Chianese et al., 2020};
\citeproc{ref-Coogan2020}{Coogan et al., 2020};
\citeproc{ref-Karchev2022}{Karchev, Coogan, et al., 2022};
\citeproc{ref-Karchev2022b}{Karchev, Anau Montel, et al., 2022};
\citeproc{ref-Mishra2022}{Mishra-Sharma \& Yang, 2022}). Second, the
differentiable nature of \texttt{caustics} allows it to be integrated
directly into machine learning workflows. This could mean using
\texttt{caustics} as part of a loss function. Alternatively, it could be
through a statistical paradigm like diffusion modelling, in which
\texttt{caustics} would be directly integrated in the sampling
procedure. It has already been shown that differentiable lensing
simulators, coupled with machine learning and diffusion modelling, can
massively improve source reconstruction in strong gravitational lenses
(\citeproc{ref-Adam2022}{Adam et al., 2022}) and in weak lensing
(\citeproc{ref-Remy2023}{Remy et al., 2023}).

\section{Conclusions}\label{conclusions}

Here we have presented \texttt{caustics}, a gravitational lensing
simulator framework which allows for greater than 100 times speedup over
traditional CPU implementations by leveraging GPU resources.
\texttt{Caustics} is fully-featured, meaning one can straightforwardly
model any strong lensing system with state-of-the-art techniques. The
code and documentation facilitate users transition from beginner to
expert by providing three interfaces which allow increasingly more
flexibility in how one wishes to model a lensing system.
\texttt{Caustics} is designed to be the gravitational lensing simulator
of the future and to meet the hundreds of thousands of lenses soon to be
discovered with modern computational resources.

\section{Acknowledgements}\label{acknowledgements}

This research was enabled by a generous donation by Eric and Wendy
Schmidt with the recommendation of the Schmidt Futures Foundation. We
acknowledge the generous software support of the University of
Washington Scientific Software Engineering Center (SSEC) at the eScience
Institute, via matching through the Schmidt Futures Virtual Institute
for Scientific Software (VISS). CS acknowledges the support of a NSERC
Postdoctoral Fellowship and a CITA National Fellowship. This research
was enabled in part by support provided by Calcul Québec and the Digital
Research Alliance of Canada. The work of A.A. and R.L. were partially
funded by NSERC CGS D scholarships. Y.H. and L.P. acknowledge support
from the National Sciences and Engineering Council of Canada grants
RGPIN-2020-05073 and 05102, the Fonds de recherche du Québec grants
2022-NC-301305 and 300397, and the Canada Research Chairs Program.
Thanks to Simon Birrer for communications regarding benchmarking of
\texttt{lenstronomy}.

\section*{References}\label{references}
\addcontentsline{toc}{section}{References}

\phantomsection\label{refs}


\begin{CSLReferences}{1}{0}
\bibitem[\citeproctext]{ref-Adam2022}
Adam, A., Coogan, A., Malkin, N., Legin, R., Perreault-Levasseur, L.,
Hezaveh, Y., \& Bengio, Y. (2022). {Posterior samples of source galaxies
in strong gravitational lenses with score-based priors}. \emph{Machine
Learning and the Physical Sciences Workshop}, E1.
\url{https://doi.org/10.48550/arXiv.2211.03812}

\bibitem[\citeproctext]{ref-Adam2023}
Adam, A., Perreault-Levasseur, L., Hezaveh, Y., \& Welling, M. (2023).
{Pixelated Reconstruction of Foreground Density and Background Surface
Brightness in Gravitational Lensing Systems Using Recurrent Inference
Machines}. \emph{951}(1), 6.
\url{https://doi.org/10.3847/1538-4357/accf84}

\bibitem[\citeproctext]{ref-hmc}
Betancourt, M. (2017). {A Conceptual Introduction to Hamiltonian Monte
Carlo}. \emph{arXiv e-Prints}, arXiv:1701.02434.
\url{https://doi.org/10.48550/arXiv.1701.02434}

\bibitem[\citeproctext]{ref-pyro}
Bingham, E., Chen, J. P., Jankowiak, M., Obermeyer, F., Pradhan, N.,
Karaletsos, T., Singh, R., Szerlip, P., Horsfall, P., \& Goodman, N. D.
(2019). Pyro: Deep universal probabilistic programming. \emph{The
Journal of Machine Learning Research}, \emph{20}(1), 973--978.

\bibitem[\citeproctext]{ref-lenstronomy}
Birrer, S., Shajib, A. J., Gilman, D., Galan, A., Aalbers, J., Million,
M., Morgan, R., Pagano, G., Park, J. W., Teodori, L., Tessore, N.,
Ueland, M., Vyvere, L. V. de, Wagner-Carena, S., Wempe, E., Yang, L.,
Ding, X., Schmidt, T., Sluse, D., \ldots{} Amara, A. (2021). Lenstronomy
II: A gravitational lensing software ecosystem. \emph{Journal of Open
Source Software}, \emph{6}(62), 3283.
\url{https://doi.org/10.21105/joss.03283}

\bibitem[\citeproctext]{ref-JAX}
Bradbury, J., Frostig, R., Hawkins, P., Johnson, M. J., Leary, C.,
Maclaurin, D., Necula, G., Paszke, A., VanderPlas, J., Wanderman-Milne,
S., \& Zhang, Q. (2018). \emph{{JAX}: Composable transformations of
{P}ython+{N}um{P}y programs} (Version 0.3.13).
\url{http://github.com/google/jax}

\bibitem[\citeproctext]{ref-Brehmer2019}
Brehmer, J., Mishra-Sharma, S., Hermans, J., Louppe, G., \& Cranmer, K.
(2019). {Mining for Dark Matter Substructure: Inferring Subhalo
Population Properties from Strong Lenses with Machine Learning}.
\emph{886}(1), 49. \url{https://doi.org/10.3847/1538-4357/ab4c41}

\bibitem[\citeproctext]{ref-Chianese2020}
Chianese, M., Coogan, A., Hofma, P., Otten, S., \& Weniger, C. (2020).
{Differentiable strong lensing: uniting gravity and neural nets through
differentiable probabilistic programming}. \emph{496}(1), 381--393.
\url{https://doi.org/10.1093/mnras/staa1477}

\bibitem[\citeproctext]{ref-Collett2015}
Collett, T. E. (2015). THE POPULATION OF GALAXY--GALAXY STRONG LENSES IN
FORTHCOMING OPTICAL IMAGING SURVEYS. \emph{The Astrophysical Journal},
\emph{811}(1), 20. \url{https://doi.org/10.1088/0004-637X/811/1/20}

\bibitem[\citeproctext]{ref-Coogan2022}
Coogan, A., Edwards, T. D. P., Chia, H. S., George, R. N., Freese, K.,
Messick, C., Setzer, C. N., Weniger, C., \& Zimmerman, A. (2022).
{Efficient gravitational wave template bank generation with
differentiable waveforms}. \emph{106}(12), 122001.
\url{https://doi.org/10.1103/PhysRevD.106.122001}

\bibitem[\citeproctext]{ref-Coogan2020}
Coogan, A., Karchev, K., \& Weniger, C. (2020). {Targeted
Likelihood-Free Inference of Dark Matter Substructure in Strongly-Lensed
Galaxies}. \emph{arXiv e-Prints}, arXiv:2010.07032.
\url{https://doi.org/10.48550/arXiv.2010.07032}

\bibitem[\citeproctext]{ref-Desdoigts2023}
Desdoigts, L., Pope, B. J. S., Dennis, J., \& Tuthill, P. G. (2023).
{Differentiable optics with $\partial$Lux: I---deep calibration of flat field and
phase retrieval with automatic differentiation}. \emph{Journal of
Astronomical Telescopes, Instruments, and Systems}, \emph{9}(2), 028007.
\url{https://doi.org/10.1117/1.JATIS.9.2.028007}

\bibitem[\citeproctext]{ref-Edwards2023}
Edwards, T. D. P., Wong, K. W. K., Lam, K. K. H., Coogan, A.,
Foreman-Mackey, D., Isi, M., \& Zimmerman, A. (2023). {ripple:
Differentiable and Hardware-Accelerated Waveforms for Gravitational Wave
Data Analysis}. \emph{arXiv e-Prints}, arXiv:2302.05329.
\url{https://doi.org/10.48550/arXiv.2302.05329}

\bibitem[\citeproctext]{ref-emcee}
Foreman-Mackey, D., Hogg, D. W., Lang, D., \& Goodman, J. (2013).
{emcee: The MCMC Hammer}. \emph{125}(925), 306.
\url{https://doi.org/10.1086/670067}

\bibitem[\citeproctext]{ref-Galan2021}
Galan, A., Peel, A., Joseph, R., Courbin, F., \& Starck, J.-L. (2021).
{SLITRONOMY: Towards a fully wavelet-based strong lensing inversion
technique}. \emph{647}, A176.
\url{https://doi.org/10.1051/0004-6361/202039363}

\bibitem[\citeproctext]{ref-GIGALens}
Gu, A., Huang, X., Sheu, W., Aldering, G., Bolton, A. S., Boone, K.,
Dey, A., Filipp, A., Jullo, E., Perlmutter, S., Rubin, D., Schlafly, E.
F., Schlegel, D. J., Shu, Y., \& Suyu, S. H. (2022). {GIGA-Lens: Fast
Bayesian Inference for Strong Gravitational Lens Modeling}.
\emph{935}(1), 49. \url{https://doi.org/10.3847/1538-4357/ac6de4}

\bibitem[\citeproctext]{ref-Hezaveh2016}
Hezaveh, Y., Dalal, N., Holder, G., Kisner, T., Kuhlen, M., \& Perreault
Levasseur, L. (2016). {Measuring the power spectrum of dark matter
substructure using strong gravitational lensing}. \emph{2016}(11), 048.
\url{https://doi.org/10.1088/1475-7516/2016/11/048}

\bibitem[\citeproctext]{ref-Hezaveh2017}
Hezaveh, Y., Perreault Levasseur, L., \& Marshall, P. J. (2017). {Fast
automated analysis of strong gravitational lenses with convolutional
neural networks}. \emph{548}(7669), 555--557.
\url{https://doi.org/10.1038/nature23463}

\bibitem[\citeproctext]{ref-Huang2021}
Huang, X., Storfer, C., Gu, A., Ravi, V., Pilon, A., Sheu, W.,
Venguswamy, R., Banka, S., Dey, A., Landriau, M., Lang, D., Meisner, A.,
Moustakas, J., Myers, A. D., Sajith, R., Schlafly, E. F., \& Schlegel,
D. J. (2021). {Discovering New Strong Gravitational Lenses in the DESI
Legacy Imaging Surveys}. \emph{909}(1), 27.
\url{https://doi.org/10.3847/1538-4357/abd62b}

\bibitem[\citeproctext]{ref-Karchev2022b}
Karchev, K., Anau Montel, N., Coogan, A., \& Weniger, C. (2022).
{Strong-Lensing Source Reconstruction with Denoising Diffusion
Restoration Models}. \emph{arXiv e-Prints}, arXiv:2211.04365.
\url{https://doi.org/10.48550/arXiv.2211.04365}

\bibitem[\citeproctext]{ref-Karchev2022}
Karchev, K., Coogan, A., \& Weniger, C. (2022). {Strong-lensing source
reconstruction with variationally optimized Gaussian processes}.
\emph{512}(1), 661--685. \url{https://doi.org/10.1093/mnras/stac311}

\bibitem[\citeproctext]{ref-Keeton2011}
Keeton, C. R. (2011). \emph{{GRAVLENS: Computational Methods for
Gravitational Lensing}}. Astrophysics Source Code Library, record
ascl:1102.003.

\bibitem[\citeproctext]{ref-Kneib2011}
Kneib, J.-P., Bonnet, H., Golse, G., Sand, D., Jullo, E., \& Marshall,
P. (2011). \emph{{LENSTOOL: A Gravitational Lensing Software for
Modeling Mass Distribution of Galaxies and Clusters (strong and weak
regime)}}. Astrophysics Source Code Library, record ascl:1102.004.

\bibitem[\citeproctext]{ref-numba}
Lam, S. K., Pitrou, A., \& Seibert, S. (2015). Numba: A llvm-based
python jit compiler. \emph{Proceedings of the Second Workshop on the
LLVM Compiler Infrastructure in HPC}, 1--6.

\bibitem[\citeproctext]{ref-GLAMER}
Metcalf, R. B., \& Petkova, M. (2014). {GLAMER - I. A code for
gravitational lensing simulations with adaptive mesh refinement}.
\emph{445}(2), 1942--1953. \url{https://doi.org/10.1093/mnras/stu1859}

\bibitem[\citeproctext]{ref-Millon2024}
Millon, M., Michalewicz, K., Dux, F., Courbin, F., \& Marshall, P. J.
(2024). {Image deconvolution and PSF reconstruction with STARRED: a
wavelet-based two-channel method optimized for light curve extraction}.
\emph{arXiv e-Prints}, arXiv:2402.08725.
\url{https://doi.org/10.48550/arXiv.2402.08725}

\bibitem[\citeproctext]{ref-Mishra2022}
Mishra-Sharma, S., \& Yang, G. (2022). {Strong Lensing Source
Reconstruction Using Continuous Neural Fields}. \emph{Machine Learning
for Astrophysics}, 34. \url{https://doi.org/10.48550/arXiv.2206.14820}

\bibitem[\citeproctext]{ref-PyAutoLens}
Nightingale, James. W., Hayes, R. G., Kelly, A., Amvrosiadis, A.,
Etherington, A., He, Q., Li, N., Cao, X., Frawley, J., Cole, S., Enia,
A., Frenk, C. S., Harvey, D. R., Li, R., Massey, R. J., Negrello, M., \&
Robertson, A. (2021). `PyAutoLens`: Open-source strong gravitational
lensing. \emph{Journal of Open Source Software}, \emph{6}(58), 2825.
\url{https://doi.org/10.21105/joss.02825}

\bibitem[\citeproctext]{ref-Nikolic2018}
Nikolic, B. (2018). {Acceleration of Non-Linear Minimisation with
PyTorch}. \emph{arXiv e-Prints}, arXiv:1805.07439.
\url{https://doi.org/10.48550/arXiv.1805.07439}

\bibitem[\citeproctext]{ref-pytorch}
Paszke, A., Gross, S., Massa, F., Lerer, A., Bradbury, J., Chanan, G.,
Killeen, T., Lin, Z., Gimelshein, N., Antiga, L., Desmaison, A., Kopf,
A., Yang, E., DeVito, Z., Raison, M., Tejani, A., Chilamkurthy, S.,
Steiner, B., Fang, L., \ldots{} Chintala, S. (2019). PyTorch: An
imperative style, high-performance deep learning library. In
\emph{Advances in neural information processing systems 32} (pp.
8024--8035). Curran Associates, Inc.
\url{http://papers.neurips.cc/paper/9015-pytorch-an-imperative-style-high-performance-deep-learning-library.pdf}

\bibitem[\citeproctext]{ref-Peng2006}
Peng, C. Y., Impey, C. D., Rix, H.-W., Kochanek, C. S., Keeton, C. R.,
Falco, E. E., Lehár, J., \& McLeod, B. A. (2006). {Probing the
Coevolution of Supermassive Black Holes and Galaxies Using
Gravitationally Lensed Quasar Hosts}. \emph{649}(2), 616--634.
\url{https://doi.org/10.1086/506266}

\bibitem[\citeproctext]{ref-Perreault2017}
Perreault Levasseur, L., Hezaveh, Y. D., \& Wechsler, R. H. (2017).
{Uncertainties in Parameters Estimated with Neural Networks: Application
to Strong Gravitational Lensing}. \emph{850}(1), L7.
\url{https://doi.org/10.3847/2041-8213/aa9704}

\bibitem[\citeproctext]{ref-Petrillo2017}
Petrillo, C. E., Tortora, C., Chatterjee, S., Vernardos, G., Koopmans,
L. V. E., Verdoes Kleijn, G., Napolitano, N. R., Covone, G., Schneider,
P., Grado, A., \& McFarland, J. (2017). {Finding strong gravitational
lenses in the Kilo Degree Survey with Convolutional Neural Networks}.
\emph{472}(1), 1129--1150. \url{https://doi.org/10.1093/mnras/stx2052}

\bibitem[\citeproctext]{ref-Remy2023}
Remy, B., Lanusse, F., Jeffrey, N., Liu, J., Starck, J.-L., Osato, K.,
\& Schrabback, T. (2023). {Probabilistic mass-mapping with neural score
estimation}. \emph{672}, A51.
\url{https://doi.org/10.1051/0004-6361/202243054}

\bibitem[\citeproctext]{ref-Rodney2021}
Rodney, S. A., Brammer, G. B., Pierel, J. D. R., Richard, J., Toft, S.,
O'Connor, K. F., Akhshik, M., \& Whitaker, K. E. (2021). {A
gravitationally lensed supernova with an observable two-decade time
delay}. \emph{Nature Astronomy}, \emph{5}, 1118--1125.
\url{https://doi.org/10.1038/s41550-021-01450-9}

\bibitem[\citeproctext]{ref-dynesty}
Speagle, J. S. (2020). {DYNESTY: a dynamic nested sampling package for
estimating Bayesian posteriors and evidences}. \emph{493}(3),
3132--3158. \url{https://doi.org/10.1093/mnras/staa278}

\bibitem[\citeproctext]{ref-Stone2023}
Stone, C. J., Courteau, S., Cuillandre, J.-C., Hezaveh, Y.,
Perreault-Levasseur, L., \& Arora, N. (2023). {ASTROPHOT: fitting
everything everywhere all at once in astronomical images}.
\emph{525}(4), 6377--6393. \url{https://doi.org/10.1093/mnras/stad2477}

\bibitem[\citeproctext]{ref-Suyu2010}
Suyu, S. H., \& Halkola, A. (2010). {The halos of satellite galaxies:
the companion of the massive elliptical lens SL2S J08544-0121}.
\emph{524}, A94. \url{https://doi.org/10.1051/0004-6361/201015481}

\bibitem[\citeproctext]{ref-Vegetti2014}
Vegetti, S., \& Vogelsberger, M. (2014). {On the density profile of dark
matter substructure in gravitational lens galaxies}. \emph{442}(4),
3598--3603. \url{https://doi.org/10.1093/mnras/stu1284}

\bibitem[\citeproctext]{ref-scipy}
Virtanen, P., Gommers, R., Oliphant, T. E., Haberland, M., Reddy, T.,
Cournapeau, D., Burovski, E., Peterson, P., Weckesser, W., Bright, J.,
van der Walt, S. J., Brett, M., Wilson, J., Millman, K. J., Mayorov, N.,
Nelson, A. R. J., Jones, E., Kern, R., Larson, E., \ldots{} SciPy 1.0
Contributors. (2020). {{SciPy} 1.0: Fundamental Algorithms for
Scientific Computing in Python}. \emph{Nature Methods}, \emph{17},
261--272. \url{https://doi.org/10.1038/s41592-019-0686-2}

\bibitem[\citeproctext]{ref-Wagner2024}
Wagner-Carena, S., Lee, J., Pennington, J., Aalbers, J., Birrer, S., \&
Wechsler, R. H. (2024). {A Strong Gravitational Lens Is Worth a Thousand
Dark Matter Halos: Inference on Small-Scale Structure Using Sequential
Methods}. \emph{arXiv e-Prints}, arXiv:2404.14487.
\url{https://doi.org/10.48550/arXiv.2404.14487}

\bibitem[\citeproctext]{ref-Welch2022}
Welch, B., Coe, D., Zackrisson, E., de Mink, S. E., Ravindranath, S.,
Anderson, J., Brammer, G., Bradley, L., Yoon, J., Kelly, P., Diego, J.
M., Windhorst, R., Zitrin, A., Dimauro, P., Jiménez-Teja, Y.,
Abdurro'uf, Nonino, M., Acebron, A., Andrade-Santos, F., \ldots{}
Vikaeus, A. (2022). {JWST Imaging of Earendel, the Extremely Magnified
Star at Redshift z = 6.2}. \emph{940}(1), L1.
\url{https://doi.org/10.3847/2041-8213/ac9d39}

\bibitem[\citeproctext]{ref-holycow}
Wong, K. C., Suyu, S. H., Chen, G. C.-F., Rusu, C. E., Million, M.,
Sluse, D., Bonvin, V., Fassnacht, C. D., Taubenberger, S., Auger, M. W.,
Birrer, S., Chan, J. H. H., Courbin, F., Hilbert, S., Tihhonova, O.,
Treu, T., Agnello, A., Ding, X., Jee, I., \ldots{} Meylan, G. (2020).
{H0LiCOW - XIII. A 2.4 per cent measurement of H\(_{0}\) from lensed
quasars: 5.3{\(\sigma\)} tension between early- and late-Universe
probes}. \emph{498}(1), 1420--1439.
\url{https://doi.org/10.1093/mnras/stz3094}

\bibitem[\citeproctext]{ref-Wong2023}
Wong, K. W. K., Isi, M., \& Edwards, T. D. P. (2023). {Fast
Gravitational-wave Parameter Estimation without Compromises}.
\emph{958}(2), 129. \url{https://doi.org/10.3847/1538-4357/acf5cd}

\end{CSLReferences}
\end{document}